\documentclass[onecolumn,english]{revtex4-2}
\usepackage[utf8]{inputenc}
\usepackage[T1]{fontenc}
\usepackage{hyperref}
\hypersetup{colorlinks=true, linkcolor=blue, filecolor=magenta, urlcolor=cyan,}
\usepackage{amsmath}
\usepackage{amsfonts}
\usepackage{amssymb}
\usepackage[version=4]{mhchem}
\usepackage{stmaryrd}
\usepackage{mathrsfs}
\usepackage{graphicx}
\usepackage[export]{adjustbox}

\begin{document}
\title{Correlation functions and properties of local distributions of frustrated phases in the ground state of a dilute Ising chain in a magnetic field}

\author{Yury Panov}

\affiliation{Ural Federal University, Institute of Natural Sciences and Mathematics,
 \\ Yekaterinburg, Russia}

\begin{abstract}
The features of the response of frustrated states to the external field are considered on the example of a diluted Ising chain. 
In the ferromagnetic case, partial ordering occurs, which leads to a decrease in entropy. 
In the antiferromagnetic case, the switching on of the field leads to the appearance of a long-range order in the system, although the state remains frustrated. 
A mapping of a one-dimensional spin model to a Markov chain is proposed, which makes it possible to study in detail the properties of correlation functions and local distributions of the states in the spin chain.

\end{abstract}

\maketitle

\section{Introduction}

The Ising model is important as a theoretical tool in the understanding of phase states and phase transitions in ferroelectric materials~\cite{Lines1977} 
and due to the concept of magnetism-driven ferroelectricity~\cite{Choi2008}. 
Unusual effects may be associated with frustrated phases, including in ferroelectrics~\cite{McQueen2008}. 
First of all, frustrated states are characterized by a nontrivial response to the external field. 
The source of frustration, in addition to the geometry of the lattice, can be impurities. 

A dilute Ising chain is the simplest model of a system where the ground state is frustrated due to the presence of impurities.
In a zero magnetic field, this model has an exact solution~\cite{Rys1969}, which is presented and analyzed in the most general form in~\cite{Balagurov1974}, and the properties of local distributions of states over the chain sites were investigated in the work~\cite{Panov2020}. 
The entropy and phase diagram of the ground state of the system in a magnetic field at a fixed impurity concentration were obtained in ~\cite{Panov2022}.
It has been shown that in a zero magnetic field, frustrated phases with a ferro- and antiferromagnetic sign of the exchange interaction constant have the same entropy at a given impurity concentration, however, with a non-zero magnetic field, the residual entropy of the frustrated ferromagnetic phase is greater than of the antiferromagnetic one.
In order to investigate the cause of this seeming contradiction, the properties of correlation functions and local distributions for frustrated phases of the ground state in a magnetic field are studied in detail in this paper.
For calculations and analysis of the results, the mapping of a one-dimensional spin model to a Markov chain is used.

The article is organized as follows. 
The section~\ref{sec:model} briefly describes the ground state phase diagram of the model.
In the section ~\ref{sec:mapping}, the principle of mapping a one-dimensional spin model to a Markov chain is formulated and a general formula for the transition matrix is obtained.
The main results of the work and their discussion are contained in the section~\ref{sec:results}. 
Brief conclusions are formulated in the section~\ref{sec:conclusion}.

\section{Model \label{sec:model}}

The Hamiltonian of a dilute Ising chain in a magnetic field can be written as follows~\cite{Shadrin2022}:
\begin{equation}
	H = -J \sum_{i} S_{z,i} S_{z,i+1} 
	+ V \sum_{i} \delta_{0,i} \delta_{0,i+1} 
	- h \sum_{i} S_{z,i} 
	- \mu \sum_{i} \delta_{0,i} .
	\label{eq:Ham}
\end{equation}
We use the pseudospin $S=1$ operator, 
where the states of the spin doublet and non-magnetic impurity correspond to $z$-projections
$S_z = \pm1$ and $S_z = 0$, 
$J$ is the exchange interaction constant, 
$V>0$ is the effective intersite interaction for impurities, 
$h$ is an external magnetic field, 
$\mu$ is a chemical potential for impurities, and 
$\delta_{0} = 1-S_z^2$ is projection operator on the impurity state. 
We use an annealed version of the model when the positions of impurities is not fixed.
It is also convenient to introduce projectors $\delta_{\pm1}$ for states with $S_z = \pm1$
and a projector for magnetic states $\delta_{s}=S_z^2=\delta_{+1}+\delta_{-1}$.

The ground state phase diagrams at a fixed concentration of impurities were studied earlier~\cite{Panov2022}. 
With a strong exchange, if $J>V>0$, either a ferromagnetic (FM) phase is realized, in which the macroscopic domains of ferromagnetically ordered and directed along the field spins are separated by domains of non-magnetic impurities, or, at $J<-V-|h|$, an antiferromagnetic (AFM) phase, which consists of macroscopic domains of antiferromagnetically ordered spins and impurity domains, and has zero magnetization. 
The FM and AFM phases have zero entropy. 
With weak exchange, if $-V-|h|<J<V$, the entropy of the system is not zero, i.e. the ground state is frustrated. 
For a weakly diluted spin chain, $0<n<1/2$, a frustrated ferromagnetic (FR-FM) or frustrated antiferromagnetic (FR-AFM) state is realized, where the FM or AFM spin clusters and single impurities alternate with each other.
The FR-FM and FR-AFM phases are separated on the phase diagram by the spin-flip field $|h|=-2J$.
At $n=1/2$, a charge-ordered (CO) state with zero entropy occurs when spin and impurity nodes alternate. 
For a highly diluted spin chain, $1/2<n<1$, the ground state is a frustrated paramagnetic (FR-PM) phase in which single spins directed along the field are separated by clusters of impurities. 

If $h=0$, the entropy of frustrated phases per site is given by the expression~\cite{Panov2020} 
\begin{equation}
	\mathcal{S} =
	- 2|m| \ln\left(2|m|\right)
	- \left(\frac{1}{2}-|m|\right) \ln\left(\frac{1}{2}-|m|\right) 
	+ \left(\frac{1}{2}+|m|\right) \ln\left(\frac{1}{2}+|m|\right) 
	+ \left(\frac{1}{2}-|m|\right) \ln2 , 
	\label{eq:s0h0-1}
\end{equation}
where $m=n-1/2$ is the deviation of the impurity concentration from the half filling.

\begin{figure}
\centering
	\includegraphics{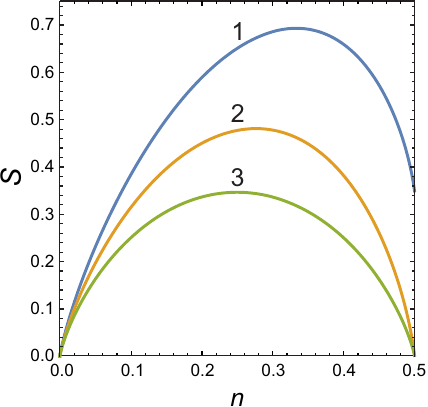}
	\caption{Entropy of the frustrated ground state phases of a dilute Ising chain in the weakly dilute case. 
	The numbers next to the curves correspond to: 
	1 --- FR-FM and FR-AFM phases at $h=0$, 
	2 --- FR-FM phase at $h\neq0$,
	3 --- FR-AFM phase at $h\neq0$.
	\label{fig:res-entr}}
\end{figure}

At $h\neq0$, the expressions for the entropy of the FR-FM and FR-AFM phases differ.
For the FR-FM phase, the  concentration dependence of entropy has the following form~\cite{Panov2022}:
\begin{equation}
	\mathcal{S} =
	- 2|m| \ln\left(2|m|\right)
	- \left(\frac{1}{2}-|m|\right) \ln\left(\frac{1}{2}-|m|\right) 
	+ \left(\frac{1}{2}+|m|\right) \ln\left(\frac{1}{2}+|m|\right) . 
	\label{eq:s0-FRFM}
\end{equation}
For a given concentration, this value is greater than entropy of the FR-AFM phase:
\begin{equation}
	\mathcal{S} =
	- |m|\ln|m| 
	- \left(\frac{1}{2}-|m|\right)\ln\left(\frac{1}{2}-|m|\right)
	- \frac{1}{2}\ln2 . 
	\label{eq:s0-FRAFM}
\end{equation}
Concentration dependences of entropy~(\ref{eq:s0h0-1}--\ref{eq:s0-FRAFM}) are shown in Fig.\ref{fig:res-entr}.
For a highly diluted chain, the concentration dependence of the entropy of the FR-PM phase is also described by the equation~\eqref{eq:s0-FRFM}, which indicates a kind of symmetry of impurity and spin states in the FM case. 
The magnetization in the FR-FM and FR-PM phases is equal to the concentration of spin centers, $M=n_s=1-n$, while in the FR-AFM phase the magnetization if defined by the impurity concentration, $M=n$.

Detailed information about the states of frustrated phases in a magnetic field can be obtained by studying correlation functions and local distributions of states on the sites of the chain.
These characteristics can be obtained by mapping a dilute Ising chain to a Markov chain.

\section{Mapping to a Markov chain ~\label{sec:mapping}}

The one-dimensional system, for which the partition function can be written using a transfer matrix, can be mapped to the Markov chain. 

For this purpose, we take as the entries of a transition matrix $\pi_{ab}$ of the Markov chain the conditional probabilities $P(b|a)$ of the state $b$ at the $(i+1)$-th site, given that the $i$-th site is in state $a$. 
Conditional probabilities are determined from the Bayes formula $P(ab)=P(a)\,P(b|a)$, where, in turn,
\begin{equation}
	P(a) = \left\langle \delta_{a,i} \right\rangle , \quad
	P(ab) = \left\langle \delta_{a,i} \, \delta_{b,i+1} \right\rangle , 
\end{equation}
and $\delta_{a,i}$ is a projector on the state $a$ for the site $i$. 
Using the transfer matrix $W$, built on the states
$a$, we find 
\begin{eqnarray}
	\left\langle \delta_{a,i} \right\rangle 
	&=& \lim_{N\to\infty} \frac{{\rm Tr} \left( W^{i-1} \delta_{a,i} W^{N-i+1} \right)}{{\rm Tr} \left( W^{N} \right)} = 
	\nonumber \\ 
	&=& \lim_{N\to\infty} 
	\frac{\sum_{k} \left\langle a|\lambda_{k}\right\rangle \lambda_{k}^{N} \left\langle\lambda_{k}|a\right\rangle}{\sum_{k} \lambda_{k}^{N}} 
	= \left\langle\lambda_1|a\right\rangle \left\langle a|\lambda_1\right\rangle , \label{eq:Pa} \\
	\left\langle \delta_{a,i} \, \delta_{b,i+1} \right\rangle 
	&=& \left\langle\lambda_1|a\right\rangle \frac{W_{ab}}{\lambda_1} \left\langle b|\lambda_1\right\rangle . 
\end{eqnarray}
Here, $\lambda_1$ is the maximum eigenvalue of the transfer matrix $W$. 
For a positive matrix, the coefficients $v_{a}=\left\langle a|\lambda_1\right\rangle$ 
can be chosen positive, according to Perron's theorem~\cite{gantmakher2000}. 
Assuming that 
\begin{equation}
	\pi_{ab} = P(b|a) 
	= \frac{\left\langle \delta_{a,i} \, \delta_{b,i+1} \right\rangle}{\left\langle \delta_{a,i} \right\rangle} , 
	\label{eq:Pab1}
\end{equation}
we find 
\begin{equation}
	\pi_{ab} = \frac{W_{ab} v_{b}}{\lambda_1 \, v_{a}} . 
	\label{eq:Pab2}
\end{equation}
The stochastic properties of the matrix $\pi$ are checked directly:
\begin{equation}
	\sum_{\beta} \pi_{ab} = \frac{1}{\lambda_1 \, v_{a}} \sum_{b} W_{ab} v_{b} = 1 . 
\end{equation}

Equation~\eqref{eq:Pab2} for obtaining a transition matrix is known in the theory of non-negative matrices~\cite{gantmakher2000}, but its physical meaning shows the equation~\eqref{eq:Pab1}. 
This allows us to use the results of a very advanced field of mathematics, the theory of Markov chains. 

The state of the system is determined by the stationary distribution $\mathbf{p}$ of the Markov chain, which can be found from the following equations 
\begin{equation}
	\sum_{a} p_{a} \pi_{ab} = p_{b} , \quad
	\sum_{a} p_{a} = 1 . 
\end{equation}
Using~\eqref{eq:Pa}, one can check that $p_{a} = P(a)$, 
and if the transfer matrix $W$ is chosen symmetric, then $p_{a}=v_{a}^{2}$. 
This also reveals the meaning of the coefficients of the maximum eigenvector of the transfer matrix. 
For the magnetization $M$ we obtain following an expression $M = \mathbf{p} \mathbf{M} $, 
where $a$th component of the vector $\mathbf{M}$ equals to the magnetization for the state $a$.

The calculation of the transition matrix $\pi$ involves finding the maximum eigenvalue $\lambda_1$ of the transfer matrix $W$, that is equivalent to calculating the partition function, so in general, all the difficulties of solving the problem of statistical description of the system are retained. 
However, the explicit form of the transition matrix $\pi$ makes it possible to efficiently find correlation functions and characteristics of local distributions, especially when considering the ground state of the system.

The pair distribution functions are related to the transfer matrix by the following equation: 
\begin{equation}
	\left\langle \delta_{a,k} \, \delta_{b,k+l} \right\rangle 
	= \left\langle\lambda_1|a\right\rangle \frac{ W_{ab}^l }{\lambda_1^l} \left\langle b|\lambda_1\right\rangle . 
\end{equation}
Using~\eqref{eq:Pab2}, we obtain 
\begin{equation}
	\left\langle \delta_{a,k} \, \delta_{b,k+l} \right\rangle 
	= p_a \, \pi^l_{ab} = \left(\pi^{tr}\right)^l_{ab} \, p_b . 
	\label{eq:ChapKol}
\end{equation}
This is a consequence of the Chapman-Kolmogorov theorem in a theory of Markov chains. 
Then one can calculate the correlation functions 
$K_{ab}(l) = \left\langle \delta_{a,k} \, \delta_{b,k+l} \right\rangle 
- \left\langle \delta_a \right\rangle \left\langle \delta_b \right\rangle$, 
and, for a dilute Ising chain, taking into account the identity 
$S_{z,k} = \delta_{{+}1,k} - \delta_{{-}1,k}$, 
spin correlation function 
$C(l) = \left\langle S_{z,k} S_{z,k+l} \right\rangle - \left\langle S_{z} \right\rangle^2$.

To consider the characteristics of local distributions of states on the chain, we denote $\sigma$ an ordered set of $k$ neighboring sites in given states: $\sigma \equiv a_{1}a_{2}\ldots a_{k}$. 
Let $(\sigma)$ be a sequence of some number of repeating blocks $\sigma$, its length is denoted by $l_{(\sigma)}$.
Probabilities $p(l_{(\sigma)})$ for a given value $l_{(\sigma)}$ obey the geometric distribution~\cite{Panov2020}:
\begin{equation}
	p(l_{(\sigma)}) = \left(1-q_{\sigma}\right) q_{\sigma}^{l-1} , 
\end{equation}
where $l$ is the number of blocks in the sequence $(\sigma)$, 
and $q_{\sigma}$ means the probability of a cycle composed of states $\sigma$, 
and this value is expressed in terms of entries of the transition matrix $\pi$:
\begin{equation}
	q_{\sigma} = \pi_{a_{1}a_{2}} \dots \pi_{a_{k-1}a_{k}} \pi_{a_{k}a_{1}} . 
	\label{eq:q-sigma}
\end{equation}
The mean length of the periodic sequence $(\sigma)$ and its variance have the form:
\begin{equation}
	\bar{l}_{(\sigma)} = \frac{1}{1-q_{\sigma}} , \quad 
	D(l_{(\sigma)}) = \frac{q_{\sigma}}{\left(1-q_{\sigma}\right)^2} . 
\end{equation}

\section{Results and discussion \label{sec:results}}

\begin{figure}
\centering
	\includegraphics{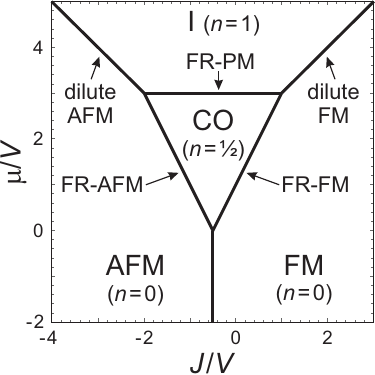}
	\caption{The phase diagrams of the 1D dilute Ising model \eqref{eq:Ham} at zero temperature 
	in plane ($J/|V|$, $\mu/|V|$) at $h=1$. 
	\label{fig:gspd}
	}
\end{figure}

The ground state phase diagram in the plane ($J/|V|$, $\mu/|V|$) can be obtained by minimizing the grand potential of a system with a Hamiltonian~\eqref{eq:Ham}. 
It is shown in Fig.\ref{fig:gspd} for $h=1$. 
Previously, a similar phase diagram~\cite{Panov2020} was obtained at $h=0$.
The impurity (I), charge-ordered, ferromagnetic, and antiferromagnetic phases 
correspond to the following configurations of the neighbouring $S_z$ states:
I $\rightarrow$ (0,0), 
CO $\rightarrow$ (0,1), 
FM $\rightarrow$ (1,1), 
AFM $\rightarrow$ (1,$-1$). 
The values of the grand potential per one site, $\omega = \Omega/N$, at zero temperature for each phase are given by the expressions:
$\omega_{I} = V - \mu$,
$\omega_{CO} = - (\mu + h)/2$, 
$\omega_{FM} = - J - h$, 
$\omega_{AFM} = J$. 
This defines the concentration of impurities $n$ in each phase: 
$n_{I} = 1$,
$n_{CO} = 1/2$, 
$n_{FM} = 0$, 
$n_{AFM} = 0$. 
Intermediate values of $n$ correspond to the states represented by the coexistence curves: 
FM/CO and AFM/CO boundaries for FR-FM and FR-AFM phases, CO/I boundary for the FR-PM phase, and FM/I and AFM/I boundaries for diluted FM and AFM phases. 

We define the transfer matrix for the model~\eqref{eq:Ham} as
\begin{equation}
W = 
\begin{pmatrix}
{\rm e}^{\beta (J+h)} & {\rm e}^{\beta (\mu+h)/2} & {\rm e}^{-\beta J} \\
{\rm e}^{\beta (\mu+h)/2} & {\rm e}^{\beta (\mu-V)} & {\rm e}^{\beta (\mu-h)/2} \\
{\rm e}^{-\beta J} & {\rm e}^{\beta (\mu-h)/2} & {\rm e}^{\beta (J-h)} 
\end{pmatrix} ,
\label{eq:Tr-matrix}
\end{equation}
where $\beta=1/T$, and we assume $k_B=1$. 

The temperature dependences of various physical quantities for this model were considered earlier in the work~\cite{Shadrin2022} using the standard formalism of a grand canonical ensemble. 
Since it is not possible to explicitly exclude the chemical potential at $h\neq0$, concentration dependences can be obtained only from the numerical solution of a system of nonlinear algebraic equations, and to study the properties of the ground state, it is also necessary to consider the limit $T\to0$. 
In this paper, we will use the mapping to the Markov chain by finding the transition matrix ~\eqref{eq:Pab2}, that is the easiest to do for the ground state. 

Note that the matrix elements~\eqref{eq:Tr-matrix} have the form $\exp\left(-\beta\omega_0\right)$, 
where $\omega_0$ is the specific grand potential of one of the phases of the ground state. 
At low temperatures, this allows us to leave in $W$ only matrix elements corresponding to this phase of the ground state, considering the remaining matrix elements as a small perturbation. 

For the undisturbed diluted phases of FM and AFM, the problem is trivial. 
Consider the corresponding transfer matrices in the limit $T\to 0$ for $h>0$:
\begin{equation}
	{\rm FM:} \quad W = 
\begin{pmatrix} 
a & 0 & 0 \\
0 & b & 0 \\
0 & 0 & 0 
\end{pmatrix} , \qquad 
	{\rm AFM:} \quad W = 
\begin{pmatrix}
0 & 0 & c \\
0 & b & 0 \\
c & 0 & 0 
\end{pmatrix} , 
\end{equation} 
where $a={\rm e}^{\beta (J+h)}$, $b={\rm e}^{\beta (\mu-V)}$, $c={\rm e}^{-\beta J}$. 
The transition matrices have the following form: 
\begin{equation}
	{\rm FM:} \quad \pi = 
\begin{pmatrix}
1 & 0 \\
0 & 1 
\end{pmatrix} , \qquad 
	{\rm AFM:} \quad \pi = 
\begin{pmatrix}
0 & 0 & 1 \\
0 & 1 & 0 \\
1 & 0 & 0 
\end{pmatrix} . 
\end{equation}
For the FM phase, the Markov chain phase space consists of 2 isolated states, 
$\Phi=\left\{{+}1,0\right\}=\left\{{+}1\right\}\bigcup\left\{0\right\}$, 
and for the AFM phase there are 2 isolated classes, 
$\Phi=\left\{{+}1,{-}1\right\}\bigcup\left\{0\right\}$, 
and the period 2 for the class $\left\{{+}1,{-}1\right\}$ corresponds to the antiferromagnetic ordering of spins in the Ising chain.
The lengts $\bar{l}_{(0)}$, 
$\bar{l}_{({+}1)}$ for the FM phase, and $\bar{l}_{({+}1{-}1)}$ for the AFM phase equal to infinity, 
this indicates the state of the chain in the form of macroscopic spin domains with the appropriate ordering,
which are separated by macroscopic domains of non-magnetic impurities.

\begin{figure}
\centering
	\includegraphics{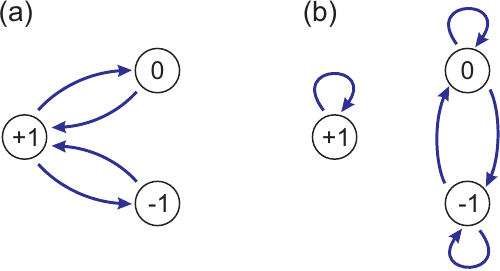}
	\caption{Transition graphs (a) of Markov chain with the transition matrix~\eqref{eq:pi-FRAFM}, 
	(b) of the 2-step Markov chain~\eqref{eq:pi-FRAFM2}. 
	\label{fig:MC-FRAFM}
	}
\end{figure}

For the FR-AFM phase at $h>0$ in the limit of $T\to0$, it is enough to consider the transfer matrix
\begin{equation}
	W = 
\begin{pmatrix}
0 & d & c \\
d & 0 & 0 \\
c & 0 & 0 
\end{pmatrix} ,
\end{equation}
where $d={\rm e}^{\beta (\mu+h)/2}$. 
For the maximum eigenvalue $\lambda_1 = \sqrt{c^2+d^2}$ the eigenvector is proportional to 
$\mathbf{v} = \left(\lambda_1,d,c\right)$, that allows us to find the transition matrix and the stationary distribution of the Markov chain:
\begin{equation}
	\pi = 
	\begin{pmatrix}
0 & \frac{d^2}{\lambda_1^2} & \frac{c^2}{\lambda_1^2} \\ 
1 & 0 & 0 \\ 
1 & 0 & 0 
\end{pmatrix} , \qquad 
	\mathbf{p} = 
	\begin{pmatrix} \frac{1}{2} \\ \frac{d^2}{2\lambda_1^2} \\ \frac{c^2}{2\lambda_1^2} \end{pmatrix} . 
\end{equation}
Since $p_0=n=1/2-|m|$ (here we took into account that $m<0$ for the FR-AFM phase), then finally we get
\begin{equation}
	\pi = 
	\begin{pmatrix}
0 & 1{-}2|m| & 2|m| \\ 
1 & 0 & 0 \\ 
1 & 0 & 0 
\end{pmatrix} , \qquad 
	\mathbf{p} = 
	\begin{pmatrix} \frac{1}{2} \\ \frac{1}{2}{-}|m| \\ |m| \end{pmatrix} . 
	\label{eq:pi-FRAFM}
\end{equation}
The graph of this Markov chain is shown in Fig.\ref{fig:MC-FRAFM}(a).

Using the Eq.~\eqref{eq:ChapKol}, we find the impurity and spin correlation functions
for the FR-AFM phase at $h\neq0$:
\begin{equation}
	K_{00}(l) = (-1)^{l}\left(\frac{1}{2} - |m|\right)^{2} , \quad
	C(l) = (-1)^{l}\left(\frac{1}{2} + |m|\right)^{2} . 
\end{equation}
Both of these functions have an infinite correlation length. 
At first glance, this sign of an ordered state contradicts the nonzero entropy of the FR-AFM phase. 
The reason for this can be understood by analyzing the properties of a Markov chain with a transition matrix~\eqref{eq:pi-FRAFM}. 
This Markov chain has a period of 2. 

Calculating a 2-step transition matrix 
\begin{equation}
	\pi^2 = 
	\begin{pmatrix}
1 & 0 & 0 \\ 
0 & 1{-}2|m| & 2|m| \\ 
0 & 1{-}2|m| & 2|m|  
\end{pmatrix} , 
	\label{eq:pi-FRAFM2}
\end{equation}
we see that for a 2-step Markov chain, the phase space contains two isolated classes, 
$\Phi=\left\{+1\right\}\bigcup\left\{0,-1\right\}$ (Fig.\ref{fig:MC-FRAFM}(b)). 
For a spin chain, this means that it is divided into two sublattices, one of which is always filled with states $+1$, i.e. completely ordered, and in the second sublattice, the states $-1$ are replaced by $0$ as the concentration of impurities increases, and therefore the magnetization is determined by unbalancing of the number of states $+1$ and $-1$, i.e. $M=n$. 
The state of the second sublattice is frustrated, its properties are determined by the corresponding block of the matrix~\eqref{eq:pi-FRAFM2}, and by the probabilities $2p_{0}=1{-}2| m|$ and $2p_{-1}=2|m|$. 
One can make sure that the spin and impurity correlation functions are zero here. 
In general, the state of the chain in the FR-AFM phase is a sequence of AFM spin clusters separated by single impurities that contain an odd number of spins and have $+1$ states at the edges.

As for the properties of local distributions, 
in the FR-AFM phase, any sequence $(\sigma)=(a)$, where $a={+}1,0,{-}1$, has a minimum length $\bar{l}_{(a)}=1$. 
For the antiferromagnetic sequence $(\sigma)=({+}1{-}1)$ 
we get $q_{{+}1{-}1}=\pi_{{+}1{-}1}\,\pi_{{-}1{+}1}=2|m|$, so the mean length 
\begin{equation}
	\bar{l}_{({+}1{-}1)} = \frac{1}{1-2|m|} 
	\label{eq:la-frafm}
\end{equation}
becomes infinite only in the absence of impurities, $m=-1/2$.

For the FR-FM phase at $h>0$ in the limit of $T\to 0$, we obtain a transfer matrix 
\begin{equation}
	W = 
\begin{pmatrix}
a & d & 0 \\
d & 0 & 0 \\
0 & 0 & 0 
\end{pmatrix} . 
\end{equation}
The maximum eigenvalue $\lambda_1 = \left(a+\sqrt{a^2+4d^2}\right)/2$ has an eigenvector that is proportional to $\mathbf{v} = \left(\lambda_1,d,0\right)$.
The phase space of the Markov chain in this case consists of 2 states, 
$\Phi=\left\{+1,0\right\}$. 
Performing calculations similar to the previous ones, we write down the transition matrix: 
\begin{equation}
	\pi = 
	\begin{pmatrix}
	\frac{4|m|}{1+2|m|} & \frac{1-2|m|}{1+2|m|} \\
	1 & 0 
	\end{pmatrix} . 
	\label{eq:P-frfm}
\end{equation}
Using Eq.~\eqref{eq:ChapKol}, we find the pair distribution functions for impurities and for spins:
\begin{eqnarray}
	\left\langle P_{0,k} P_{0,k+l} \right\rangle 
	&=& \left(\frac{1}{2} - |m|\right)^{2} 
	+ (-1)^l \left(\frac{1}{4}-m^2\right) \left(\frac{1-2|m|}{1+2|m|}\right)^{l} , \\ 
	\left\langle S_{z,k} S_{z,k+l} \right\rangle 
	&=& \left(\frac{1}{2} + |m|\right)^{2} 
	+ (-1)^l \left(\frac{1}{4}-m^2\right) \left(\frac{1-2|m|}{1+2|m|}\right)^{l} . 
\end{eqnarray}
The first terms in these expressions are equal to 
$\left\langle \delta_{0} \right\rangle^{2}$ and $\left\langle S_{z} \right\rangle^{2}$, 
so we get 
\begin{equation}
	K_{00}(l) = C(l) = (-1)^l \left(\frac{1}{4}-m^2\right) e^{-\frac{l}{\xi}} , \quad
	\xi = \left[\ln \left(\frac{1+2|m|}{1-2|m|}\right) \right]^{-1} . 
	\label{eq:xi-FRFM} 
\end{equation}
For the FR-FM phase, the correlation length $\xi$ becomes infinite only at $m\to0$, i.e. when the chain is half-filled with impurities. 
Note that in the limit $m=-1/2$, the correlation length turns to zero, since there are no fluctuations in the pure Ising chain at $h\neq0$ and $T=0$. 

Next, we find in the FR-FM phase at $h>0$ the mean length of the spin sequence, which in this case has the form $(\sigma)=({+}1)$. 

Using the Eqs~\eqref{eq:q-sigma} and~\eqref{eq:P-frfm}, we get
\begin{equation}
	q_{{+}1} = \frac{4|m|}{1+2|m|} , \quad
	\bar{l}_{({+}1)} = \frac{1+2|m|}{1-2|m|} . 
	\label{eq:ls-frfm}
\end{equation}
We note, that $\bar{l}_{({+}1)}$ is maximal at $m=-1/2$ 
and reaches a minimum value $\bar{l}_{({+}1)} = 1$ at a half-filling , $m=0$. 
If $m\simeq -1/2$, then $2\bar{l}_{({+}1)}$ is close to $\bar{l}_{({+}1{-}1)}$ in the FR-AFM phase. 
For the sequence of impurities we have $\bar{l}_{(0)} = 1$, which corresponds to isolated impurities. 
For a charge-ordered sequence $(\sigma)=({+}10)$ we get 
\begin{equation}
	\bar{l}_{({+}10)} = \frac{1+2|m|}{4|m|} . 
	\label{eq:l01-frfm}
\end{equation}
As one can see, at a half-filling $\bar{l}_{({+}10)}$ becomes infinite.

For the FR-PM phase at $h>0$ from the transfer matrix  
\begin{equation}
	W = 
\begin{pmatrix}
0 & d & 0 \\
d & b & 0 \\
0 & 0 & 0 
\end{pmatrix} ,
\end{equation}
we obtain that the phase space of the Markov chain is 
$\Phi=\left\{+1,0\right\}$, and the transition matrix has the form 
\begin{equation}
	\pi = 
	\begin{pmatrix}
	 0 & 1 \\ 
	 \frac{1-2|m|}{1+2|m|} & \frac{4|m|}{1+2|m|} 
	\end{pmatrix} . 
\end{equation}
Compared to the FR-FM phase, in the FR-PM phase at $h>0$ and $m>0$, the states $0$ and ${+}1$ are reversed: 
in this case, the individual spins directed along the field are separated by sequences of impurities with the  mean length $\bar{l}_{(0)} = \frac{1+2|m|}{1-2|m|}$. 
As a result, the expression ~\eqref{eq:l01-frfm} for $\bar{l}_{({+}10)}$ remains the same, 
and the expressions for the correlation functions of the FR-PM and FR-FM phases and the concentration dependences of entropy have the same form.

\begin{figure}
\centering
	\includegraphics{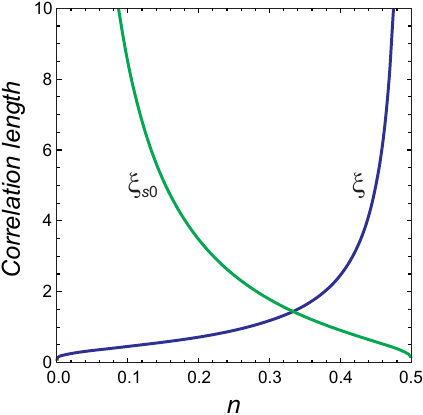}
	\caption{Dependences \eqref{eq:xi-FRFM} and \eqref{eq:xis0} of the spin correlation length 
	at $h\neq0$ and $h=0$ on the concentration of impurities $n$.
	\label{fig:corr-lengths}
	}
\end{figure}

Now we compare the obtained results with the case $h=0$, when the transition matrices $\pi$ for FR-FM, FR-AFM, and FR-PM phases have the following form: 
\begin{equation}
	\begin{pmatrix}
	1{-}\alpha & \alpha & 0 \\
	\frac{1}{2} & 0 & \frac{1}{2} \\
	0 & \alpha & 1{-}\alpha 
	\end{pmatrix} , \quad 
	\begin{pmatrix}
	0 & \alpha & 1{-}\alpha \\
	\frac{1}{2} & 0 & \frac{1}{2} \\
	1{-}\alpha & \alpha & 0 
	\end{pmatrix} , \quad 
	\begin{pmatrix}
	0 & 1 & 0 \\
	\frac{\alpha}{2} & 1{-}\alpha & \frac{\alpha}{2} \\
	0 & 1 & 0 
	\end{pmatrix} , 
\end{equation}
where $\alpha=(1-2|m|)/(1+2|m|)$. 
Using equation~\eqref{eq:ChapKol}, we obtain that the impurity correlation functions for FR-FM, FR-AFM and FR-PM phases are the same and are described by the same expression~\eqref{eq:xi-FRFM} as for $h\neq0$.
The spin correlation length $\xi_{s0}$ for FR-FM and FR-AFM phases is also finite: 
\begin{equation}
	C_{0}(l) = (\pm1)^{l} \left(\frac{1}{2} + |m|\right) e^{-\frac{l}{\xi_{s0}}} , \quad 
	\xi_{s0} = \left[\ln \left(\frac{1+2|m|}{4|m|}\right) \right]^{-1} , 
	\label{eq:xis0} 
\end{equation}
where the upper and lower signs correspond to the cases $J>0$ and $J<0$. 
Fig.\ref{fig:corr-lengths} shows the dependence of the spin correlation length at $h\neq0$ and $h=0$ on the concentration of impurities $n=1/2+m$. 
In the FR-PM phase, $C_{0}(l)=0$. 

Thus, the FR-FM and FR-AFM phases show a fundamentally different response to the turning on of an external magnetic field. 

For the FR-FM phase at $h\neq0$, the spin correlation length decreases relative to the value at $h=0$ if $0<n<1/3$ and increases if $1/3<n<1/2$, and the impurity correlation length does not change. 
At the same time, the average lengths of the spin sequences $(\sigma)=({+}1)$ and $(\sigma)=({-}1)$ for $h=0$ coincide with the average length of the sequence $(\sigma)=({+}1)$ for $h\neq0$. 
This corresponds to the reversal of finite FM clusters in the direction of the field at $h\neq0$. 

In the FR-AFM phase, the spin and impurity correlation lengths are finite at $h=0$, but become infinite at $h\neq0$. 
The transformation of the state occurring in this case is significant: one sublattice turns out to be completely filled with spins directed along the field, and oppositely directed spins and impurities are located only on the other sublattice. 
At $h\neq0$, pairs of $0s$ disappear, where $s$ are spins directed opposite the field. 
If at $h=0$ the number of sites in the AFM cluster is any, then at $h\neq0$ this number can only be odd, and at the edges of the cluster the spins should be directed along the field. 
If $h\neq0$, the FR-AFM phase is more ordered at a given impurity concentration than the FR-FM phase and has a lower entropy.

It is worth to note, that that if we combine the states ${+}1$ and ${-}1$ into one spin state $s$ using the projector $\delta_{s}$, the picture will change. 
The transition matrix on phase space $\Phi=\left\{s,0\right\}$ for FR-FM and FR-AFM phases at $h=0$ and $h\neq0$ will have the form~\eqref{eq:P-frfm}. 
On the one hand, it shows that the mean length of the spin sequence $(\sigma)=(s)$ remains unchanged in all cases. 
On the other hand, when using this technique, the opportunity to study the subtle effects of ordering in the magnetic field for the FR-AFM phase is lost.

\section{Conclusion \label{sec:conclusion}}
In this paper, we have proposed a method for mapping one-dimensional spin models to a Markov chain and analyzed 
correlation functions and properties of local distributions of states in the ground state of a dilute Ising chain. 
It is shown that although at zero magnetic field the states of frustrated phases of a weakly diluted chain are equivalent, their changes when the field is switched on differ fundamentally depending on the type of magnetic ordering. 
For the ferromagnetic case, FM clusters of finite size are ordered in the direction of the field. 
In the frustrated antiferromagnetic phase, when the field is turned on, a long-range order occurs due to filling one sublattice with spins directed along the field. 
At the same time, the state of the system remains frustrated due to the macroscopic number of configurations corresponding to the distribution of impurities and spins directed opposite the field in another sublattice. 

\bigskip
\section*{Funding}

The work was supported under grant FEUZ-2023-0017 of the Ministry of Science and Higher Education of the Russian Federation.


\end{document}